\documentclass[fleqn,usenatbib]{mnras}

\usepackage{newtxtext,newtxmath}

\usepackage[T1]{fontenc}
\usepackage{booktabs}
\DeclareRobustCommand{\VAN}[3]{#2}
\let\VANthebibliography\thebibliography
\def\thebibliography{\DeclareRobustCommand{\VAN}[3]{##3}\VANthebibliography}

\usepackage{graphicx}	
\usepackage{amsmath}	

\newcommand{\hMpc}{{\ifmmode{\;h^{-1}{\rm Mpc}}\else{$h^{-1}$Mpc}\fi}}
\newcommand{\hkpc}{{\ifmmode{\;h^{-1}{\rm kpc}}\else{$h^{-1}$kpc}\fi}}
\newcommand{\hMsun}{{\ifmmode{\;h^{-1}{\rm {M_{\odot}}}}\else{$h^{-1}{\rm{M_{\odot}}}$}\fi}}

\newcommand{\Mstar}{{\ifmmode{\;M_{*}}\else{$M_{*}$}\fi}}
\newcommand{\Mhalo}{{\ifmmode{\,M_{\rm halo}}\else{$M_{\rm halo}$}\fi}}
\newcommand{\ltsima}{$\; \buildrel < \over \sim \;$}
\newcommand{\gtsima}{$\; \buildrel > \over \sim \;$}
\newcommand{\lsim}{\lower.5ex\hbox{\ltsima}}
\newcommand{\gsim}{\lower.5ex\hbox{\gtsima}}

\newcommand{\thethree}{{\sc The300}}
\newcommand{\theth}{{\sc The Three Hundred}}
\newcommand{\ahf}{\textsc{Amiga's Halo Finder}}

\newcommand{\simba}{\textsc{Gizmo-SIMBA}}
\newcommand{\gadgetx}{\textsc{Gadget-X}}
\newcommand{\gadgetmusic}{\textsc{Gadget-MUSIC}}

\title[AI-assisted cluster mass distribution]{The Three Hundred Project: Mapping The Matter Distribution in Galaxy Clusters Via Deep Learning from Multiview Simulated Observations}

\author[Daniel de Andres et.al.]{
\parbox{\textwidth}
{Daniel de Andres$^{1,2}$\thanks{daniel.deandres@uam.es},
Weiguang Cui$^{3,1,2}$\thanks{Talento-CM fellow}, 
Gustavo Yepes$^{1,2}$, 
Marco De Petris$^{4}$,
Antonio Ferragamo$^{4,5}$,
Federico De Luca$^{4}$,
Gianmarco Aversano$^{6}$ and
Douglas Rennehan$^{7}$
}
\\
\\
$^{1}$ Departamento de Física Teórica, M-8, Universidad Autónoma de
Madrid, Cantoblanco 28049, Madrid, Spain\\
$^{2}$ Centro de Investigación Avanzada en Física Fundamental,(CIAFF), Universidad Aut\'{o}noma de Madrid, Cantoblanco, 28049 Madrid, Spain\\
$^{3}$ Institute for Astronomy, University of Edinburgh, Royal
Observatory, Edinburgh EH9 3HJ, UK \\
$^{4}$ Dipartimento di Fisica, Sapienza Universitá di Roma, Piazzale Aldo Moro, 5-00185 Roma, Italy \\
$^{5}$ Instituto de Astrofísica de Canarias (IAC) La Laguna, 38205, Spain \\
$^{6}$ EURANOVA, Mont-Saint-Guibert, Belgium\\
$^{7}$ Center for Computational Astrophysics, Flatiron Institute, 162 Fifth Avenue, New York, NY, 10010, USA
}

\date{Accepted XXX. Received YYY; in original form ZZZ}

\pubyear{2015}

\begin{document}
\label{firstpage}
\pagerange{\pageref{firstpage}--\pageref{lastpage}}
\maketitle

\begin{abstract}
A galaxy cluster as the most massive gravitationally-bound object in the Universe, is dominated by Dark Matter, which unfortunately can only be investigated through its interaction with the luminous baryons with some simplified assumptions that introduce an un-preferred bias. In this work, we, {\it for the first time}, propose a deep learning method based on the U-Net architecture, to directly infer the projected total mass density map from idealised observations of simulated galaxy clusters at multi-wavelengths.  The model is trained with a large dataset of simulated images from clusters of {\sc The Three Hundred Project}. Although Machine Learning (ML) models do not depend on the assumptions of the dynamics of the intra-cluster medium, our whole method relies on the choice of the physics implemented in the hydrodynamic simulations, which is a limitation of the method. Through different metrics to assess the fidelity of the inferred density map, we show that the predicted total mass distribution is in very good agreement with the true simulated cluster. Therefore, it is not surprising to see the integrated halo mass is almost unbiased, around 1 per cent for the best result from multiview, and the scatter is also very small, basically within 3 per cent. This result suggests that this ML method provides an alternative and more accessible approach to reconstructing the overall matter distribution in galaxy clusters, which can complement the lensing method.

\end{abstract}




\section{Introduction}
Estimating the matter content in galaxy clusters \citep[see][for a review]{Kravtsov2012} is crucial for cosmological studies due to the fact that they are the biggest gravitationally bound objects originating from small density fluctuations in the early universe. Thus cosmological parameters can be constrained by studying the abundance of galaxy clusters as a function of the mass and redshift \citep[e.g.][]{Allen2011, Planck2016:Clusters, Pratt2019, Salvati2022}. Galaxy clusters are mainly composed of Dark Matter (DM), which is about 80\% of their total mass, diffused hot gas (about 12\%), i.e., intra-cluster medium (ICM), and stars, mainly in galaxies (the remaining $8\%$). 

Galaxies within clusters are normally observed at different optical bands. Both ground-based and space-based instruments have been used to measure these galaxy properties through their spectrum energy distributions. 
For example, the Sloan Digital Sky telescope (SDSS\footnote{\url{https://www.sdss.org}}) and the Hubble Space Telescope (HST\footnote{\url{https://science.nasa.gov/mission/hubble}})  have been crucial for the studies of galaxy clusters as well as cosmology. The recent photometric surveys, e.g., the Dark Energy Survey (DES\footnote{\url{https://www.darkenergysurvey.org}}) using the Dark Energy Camera and the Javalambre Physics of the Accelerating Universe Astrophysical Survey J-PAS\footnote{\url{https://www.j-pas.org/survey}} from the Javalambre Survey Telescope, provide an unprecedented amount of data for understanding our Universe. Not to mention the recently launched space telescopes James Webb Space Telescope (JWST\footnote{\url{https://www.jwst.nasa.gov}}) and Euclid\footnote{\url{https://www.esa.int/Science_Exploration/Space_Science/Euclid}}. While optical surveys explore galaxy members, the ICM is targeted through X-ray emission and the inverse-Compton scattering of the cosmic microwave background (CMB) photons, i.e. the Sunyaev-Zel'dovich (SZ) effect \citep{sunyaev1972}. The SZ effect can be observed at millimetre wavelengths and different instruments have managed to detect more than a thousand clusters through it, such as the Planck satellite \citep[PSZ2;][]{PlanckPSZ2}, the Atacama Cosmology Telescope \citep[ACT;][]{ACTcatalog} and the South Pole Telescope \citep[SPT;][]{SPTcatalog}. Nevertheless, Planck is the only all-sky survey for SZ clusters. On the contrary, clusters X-ray observations are more recently explored by eROSITA \citep{erositasurvey} and targeted XMM-Newton observations, such as the  CHEX-MATE project \citep{chexmatesurvey}.

Although the total mass of a cluster of galaxies is not observable, it can be inferred by observing its baryonic components. It can be estimated from the dynamics \citep[e.g.][]{Biviano2006} or the abundance/richness \citep[e.g.][]{Rozo2009} of the member galaxies, or from the ICM radial profiles in X-ray and SZ observations with the assumption of the diffused gas is distributed following the hydrostatic equilibrium (HE) hypothesis \citep[e.g.][]{Gianfagna2021,Gianfagna2023}. Alternatively, their physical quantities, which are integrated inside a fixed aperture in the sky, strictly related to the mass of the object can be selected under the self-similarity assumption \citep{bryan1998} as suitable observational proxies. However, these estimated masses rely upon theoretical assumptions, such as HE, and are therefore possibly affected by bias. 
Nevertheless, only weak gravitational lensing (WL) is sensitive to all matter along the line of sight and measures the total projected matter \citep{Becker2011,Herbonnet2022}. However, gravitational lensing observations are not numerous, with tens of images, in comparison to the hundreds or few thousands of observations of galaxy clusters available for SZ, X-ray and stars. In addition, mass inference from WL also suffers from systematics due to the theoretical assumptions when reconstructing mass (kappa) maps from shear \citep{Kaiser1993:WLclassical}, similar to the inference of the mass from SZ or X-ray. Recent data-driven approaches based on simulations and convolutional neural networks have proven to outperform conventional methods \citep{Hong2021:CNNWL} also for the task of WL Mass Reconstruction. 

Machine Learning (ML) has been applied in astronomy since quite a long ago \citep[e.g][]{Odewahn1992}, but only in past years we have witnessed an unprecedented increase of Deep Learning (DL) methods \citep[for a review, see e.g.][]{Huertas2022,exmachina2023}, which implies a change in the paradigm from applying specific algorithms to fully data-driven science. We stress here that applying deep learning in real observations of galaxy clusters is inherently more challenging than in simulations and accordingly most of the literature is limited to only utilising simulated data. For example, DL has also been used for identifying galaxy cluster members from HST images \citep{Angora2020:GalaxyMembersHST}, increasing the resolution (super-resolution) and de-noising XMM-Newton images \citep{Sweere2022:super-resolution}, and deprojecting and deconvolving galaxy cluster X-ray temperature profiles within the CHEX-MATE collaboration \citep{Iqbal2023}. In our interests, recent studies have shown that galaxy cluster masses can be estimated using deep learning methods from catalogues of simulated galaxy clusters \citep{Ntampaka2019,Ho2019,Gupta2020,Kodi2020,Yan2020,Gupta2021,ho2021,deAndres2022Planck,Ferragamo2023,Ho2023}, which outperform these classical methods. Fortunately, only recently these methods have also been applied to infer the mass of galaxy clusters of real surveys at different wavelengths, \cite{Kodi2021} applied deep learning to infer galaxy clusters masses from the SDSS Legacy Survey \citep{SDSSsurvey}, \cite{Ho2022Coma} estimated the dynamical mass of the Coma cluster, \cite{deAndres2022Planck} inferred the masses of the full-sky Planck satellite PSZ2 catalogue, \cite{Krippendorf2023} followed a machine learning approach to infer galaxy cluster masses from eROSITA X-ray images.

Notwithstanding, previous works on mass inference focus on the mass inside a sphere of a certain radius (i.e. $R_{200}$\footnote{$R_{200}$ stands for the radius of the sphere at which the density is 200 times the critical density of the Universe at the corresponding redshift} or $R_{500}$). Recently, \cite{Ferragamo2023} pushed these ML applications even further by showing that ML can predict the mass radial profile using SZ idealised simulated images and thus, being able to theoretically obtain valuable information on internal properties of galaxy clusters as the concentration. As a general remark, all these studies suggest that the estimated masses are not affected by observational biases and assumptions due to the fact that they do not rely on hypotheses about the dynamics nor the hydrostatic equilibrium with spherical symmetry of the ICM, but rather on the quality of dataset provided by cosmological simulations.

In this work, for the first time, we explore Deep Learning models to infer the projected matter density fields from simulated baryonic observations. To this end, we utilise the {\sc{The Three Hundred}} \citep[\thethree;][]{Cui2018}: a set of ``zoom-in'' hydrodynamical and cosmological simulations to generate the input idealised observations: SZ, X-ray and star maps, and the output maps, mass maps. The dataset created for this purpose corresponds to a set of idealised observations free from the observational impacts typical of real instruments, such as noise and point sources. Therefore, this work represents the needed theoretical proof-of-concept study where the limitations of deep learning are tested. For further details regarding our dataset, we refer the reader to \autoref{sec-2}. 
 
The manuscript is organised as follows: in \autoref{sec-2} the creation of the dataset is discussed, {\theth{}} set of hydrodynamic simulations and the particular selection of halo-sized objects that are optimal for training ML models and the characteristics of our input maps, Compton-$y$ parameters maps (SZ), X-ray surface brightness (X-ray) and star density maps, and output mass density maps. In \autoref{sec-3} we discuss the Deep learning model and architectures considered for this project as well as how the models are trained. In \autoref{sec-4}, the results of the model applied to the test set are shown. To this end, we design a set of test metrics to assess the fidelity of the predicted mass density maps: pixel-wise statistics, cylindrical radial profiles, power spectra and Maximum Mean Discrepancy. Finally, in \autoref{sec-5} the main conclusions of the work are drawn.

\section{Dataset}\label{sec-2}

\subsection{The Three Hundred Simulations}
In order to create the dataset used for training our DL model, we use {\sc The300}\footnote{\url{https://the300-project.org}} simulations \citep{Cui2018}. {\sc{The300}} simulation project focuses on a set of 324 ``zoom-in'' hydrodynamic simulations centred on the most massive halos of the  {\sc{MultiDark}} simulation \citep[{\sc{MDPL2}};][]{MDPL2} utilising the cosmological parameters inferred by Planck \citep{Planckparameters}. The full MDPL2 simulation contains $3840^3$ particles whose mass is $1.5\times 10^{9}\hMsun$. Moreover, each cluster region covers 15\hMpc\ and they have been simulated with different baryonic models:  
\gadgetmusic{} \citep{Sembolini2013}, \gadgetx{} \citep{Murante2010, Rasia2015}, \simba{}, \citet{Dave2019,Cui2022}). However, in this work, we only make use of \gadgetx{} \citep{Cui2018} and \simba{} \cite{Cui2022} runs. 

In {\thethree} simulations halos are identified by the \ahf{} package \citep[AHF;][]{AHF} and only halos with $M_{200} > 10^{13.5} h^{-1} M_\odot$, at redshift $z\simeq 0$ and free from contamination are selected. Note that $M_{200}$ stands for the mass inside a sphere of density 200 times the critical density of the Universe at the corresponding redshift. Bear in mind that for {\thethree} simulations, the scaling relations do not depend on redshift at least up to $z\simeq 1$ \citep{deandres2022baryon}. Free from contamination means that heavy (or low resolution) dark-matter particles entering from outside the re-simulated volume are not present inside our selected halos. The selection is first done for \gadgetx, the counterpart halos from \simba\ are selected to have the same number of objects and mass distribution as in \gadgetx{}.

\begin{figure}
\includegraphics[width=\columnwidth]{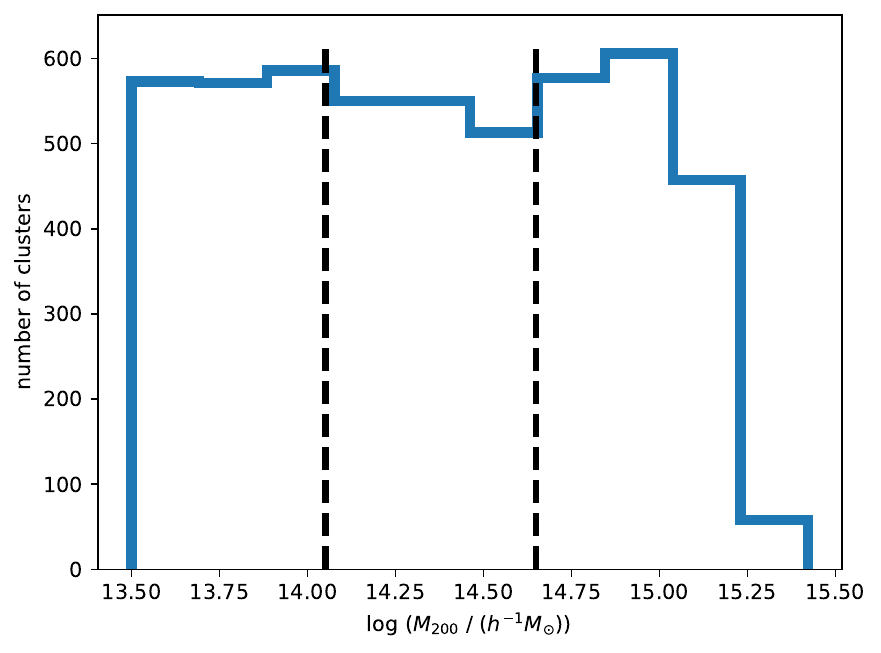}
\caption{Mass distribution of the selected galaxy clusters sample. Note that the selected objects follow an approximately flat distribution in the logarithm of the mass. Dashed black vertical lines correspond to mass thresholds that divide the sample in 3 mass bins containing an equal number of clusters and they are at $\log (M_{200}/(h^{-1}M_{\odot}))=14.05$ and $\log (M_{200}/(h^{-1}M_{\odot}))=14.65$.}
\label{fig:massstats}
\end{figure}

Furthermore, to increase the statistics of underrepresented clusters we added objects from other snapshots ($z = 0.022$, $0.045$, $0.069$, $0.093$, $0.117$) to account for a flat mass distribution as in \cite{Ferragamo2023}, which is presented in \autoref{fig:massstats}. In that figure, the number of galaxy clusters as a function of mass is shown, and the whole sample is equally split into 3 bins with two vertical lines. Moreover, the {\sc The300} simulations contain a rich variety of massive galaxy clusters with different dynamical states \citep{DeLuca2021}. Note that at the end of the selection procedure, 2518 halos are selected from \gadgetx{} and 2523 from \simba{}, with the same flat distribution in mass. We refer the reader to the Appendix A for more information.

\subsection{Simulated multiview images}
From the selected halos, we further compute the Compton-$y$ parameter -- a dimensionless measure of the
SZ effect, X-ray surface brightness, star mass density and total mass density maps. Per each halo in our dataset, 29 different line-of-sight projections are considered to create 146,189 maps in total, a large enough dataset for training our ML model. To test the effectiveness of machine learning models and avoid observational biases, the generated images correspond to a set of idealised maps in which the angular resolution is sufficiently high for all of them regardless of their masses. As such, we have created a dataset in which the size of each map is $2\times R_{200}$, all sampled with the same number of pixels $N_{\text{pix}}=640$ and therefore all maps are equally resolved.

Although all maps are initially created with a resolution of $640\times 640$ pixels, they are smoothed with a Gaussian kernel of $FWHM\simeq0.015 R_{200}$ and they are re-gridded to a lower resolution of $80\times 80$ pixels, which is needed for computational efficiency. Nevertheless, these ``low-resolution'' final maps are still very well resolved, and the angular resolution for all maps is below 1 arcmin. The idea that we desire to convey is that this dataset corresponds to an ideal theoretical dataset from hydro-simulations which can be easily applied to the ML models for testing their performance. In detail, the SZ maps, X-ray surface brightness maps, star mass density maps and total mass density maps are computed as follows:

\textbf{Compton-y parameter maps} ($y$) correspond to the integrated pressure along the observer's line of sight (l.o.s.) $dl$.
\begin{equation}\label{eq:defy}
    y = \frac{\sigma_{\text{T}}k_{\text{B}}}{m_{\text{e}}c^{2}}\int n_{\text{e}}T_{\text{e}}dl \text{ ,}
\end{equation}
where $\sigma_{\text{T}}$ is the Thomson cross section, $k_{\text{B}}$ is the Boltzmann constant, $c$ the speed of light, $m_{\text{e}}$ the electron rest-mass, $n_{\text{e}}$ the electron number density and $T_{\text{e}}$ the electron temperature. These maps are computed using the publicly available package {\sc PYMSZ}\footnote{\url{https://github.com/weiguangcui/pymsz}} \citep{Cui2018}.

\textbf{X-ray surface brightness maps} (X-ray) are estimated by computing the X-ray emission by thermal bremsstrahlung in the hot intracluster medium using a wrapper of PyAtomDB\footnote{\url{https://atomdb.readthedocs.io/en/master/}} package to compute X-ray luminosity\footnote{\url{https://github.com/rennehan/xraylum}}.  First, using the atomic database, we estimated the bolometric X-ray luminosity of gas particles in the simulation. Then we projected the sum of the luminosity values along the observer's l.o.s. The projected luminosity $L_{ij}$, is then divided by the surface area $\Sigma$ of the pixel $i,j$:
\begin{equation}
    \text{X-ray}_{ij}= L_{ij}/\Sigma\text{ ,}
\end{equation}
which is
\begin{equation}
    \Sigma = (R_{200}/40)^{2}\text{ .}
\end{equation}

\textbf{Stellar density maps} are generated by projecting the sum of the masses of the star particles $\sum_{\text{p}} M_{*\text{p}}$ along the observer's line of sight, that represent the pixel $(i,j)$. This value is then divided by the surface area of the pixel $\Sigma$:
\begin{equation}
    \text{star}_{i,j} = \sum_{p}M_{*,\text{p}}/\Sigma\text{ .}
\end{equation}
We note that opting for stellar density as the stars simulated maps requires some clarifications: 1) Firstly, stellar observations in surveys are all presented as multi-band observations, e.g., SDSS\footnote{\url{https://www.sdss.org}}. As a first proof-of-concept approximation, the density field of the stars reflects their spatial distribution and thus, for this purpose, as shown in \cite{Yan2020} $M_{500}$ could be inferred from star density maps. 2) We can, for sure, compute luminosity maps from stars in the simulation by using Stellar Population Synthesis models, such as SPSM \citep{Devriendt1999:SPSM,Bruzual2003:SPSM}. 
This approach is closer to the real observation images.
However, we argue that (a) the mass density map from observations can be derived by fitting the different colour bands to a Stellar Energy Distribution, using the same SPSM, albeit a little bit higher uncertainties and complexities due to several hypotheses on the metallicity, stellar mass function, etc... (b) there should be not much difference in practice with either luminosity or stellar mass maps. This is because the galaxies' luminosity-to-mass ratio is approximately constant for The300 simulation, see figure 8 of \cite{Cui2018}. (c) to apply our ML models to real observation images, there are still multiple steps to take care of, such as instruments, and background noise. Therefore, we adopt the stellar mass maps for simplicity in this concept-proofing paper and leave the proper reproduction of mock observation images in the following work.

\textbf{Mass density maps} are generated by projecting the sum of the masses of all particles, i.e. gas, star, dark matter and black hole particles in the observer's line of sight. Similarly, this value is divided by the surface area of a pixel. 

\begin{figure*}
\includegraphics[width=1.0\textwidth]{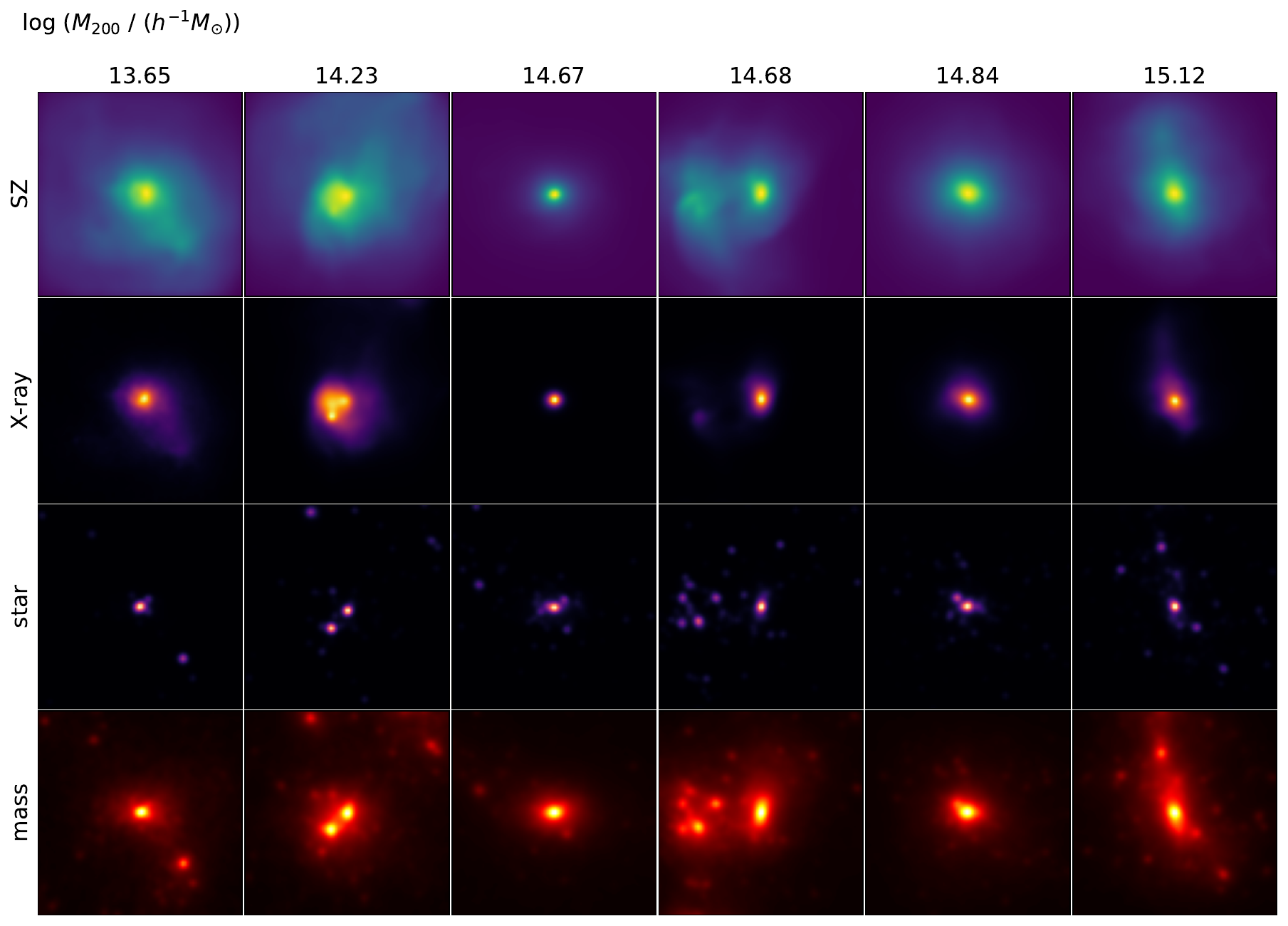}
\caption{From top to bottom, we show some examples of maps used for training the U-Net model: SZ, X-ray, star and mass. The examples' mass $M_{200}$ is shown at the top of each column. The size of all maps is $2\times R_{200}$ regardless of the mass.}
\label{fig:maps}
\end{figure*}

In \autoref{fig:maps}, examples of our simulated maps are displayed for different cluster masses. With Compton-y maps, X-ray surface brightness maps and star density maps, we aim to infer total mass density maps shown at the bottom by using a deep learning model that is able to capture the non-linear relations between input (SZ, X-ray, star) and output (total mass) maps. This model can end-to-end translate one input map into mass density. The choice of the particular model used and the training of that model are described in the next section. 

\begin{table}
\begin{tabular}{@{}ccc@{}}
\toprule
Map                      & Short name & Units                         \\ \midrule
Compton-y parameter      & SZ         &           \\
Bolometric X-ray surface brightness & X-ray      & $\text{erg}\text{ s}^{-1}\text{kpc}^{-2}$ \\
Star density maps        & star       & $\hMsun \text{ kpc}^{-2}$    \\
Mass density maps        & mass       & $\hMsun \text{ kpc}^{-2}$    \\ \bottomrule
\end{tabular}\caption{Dataset of simulated maps from {\theth{}} simulations.}
\end{table}

\section{Methods}\label{sec-3}
\subsection{The Deep Learning Model}\label{sec:Model}

\begin{figure*}
\includegraphics[width=0.9\textwidth]{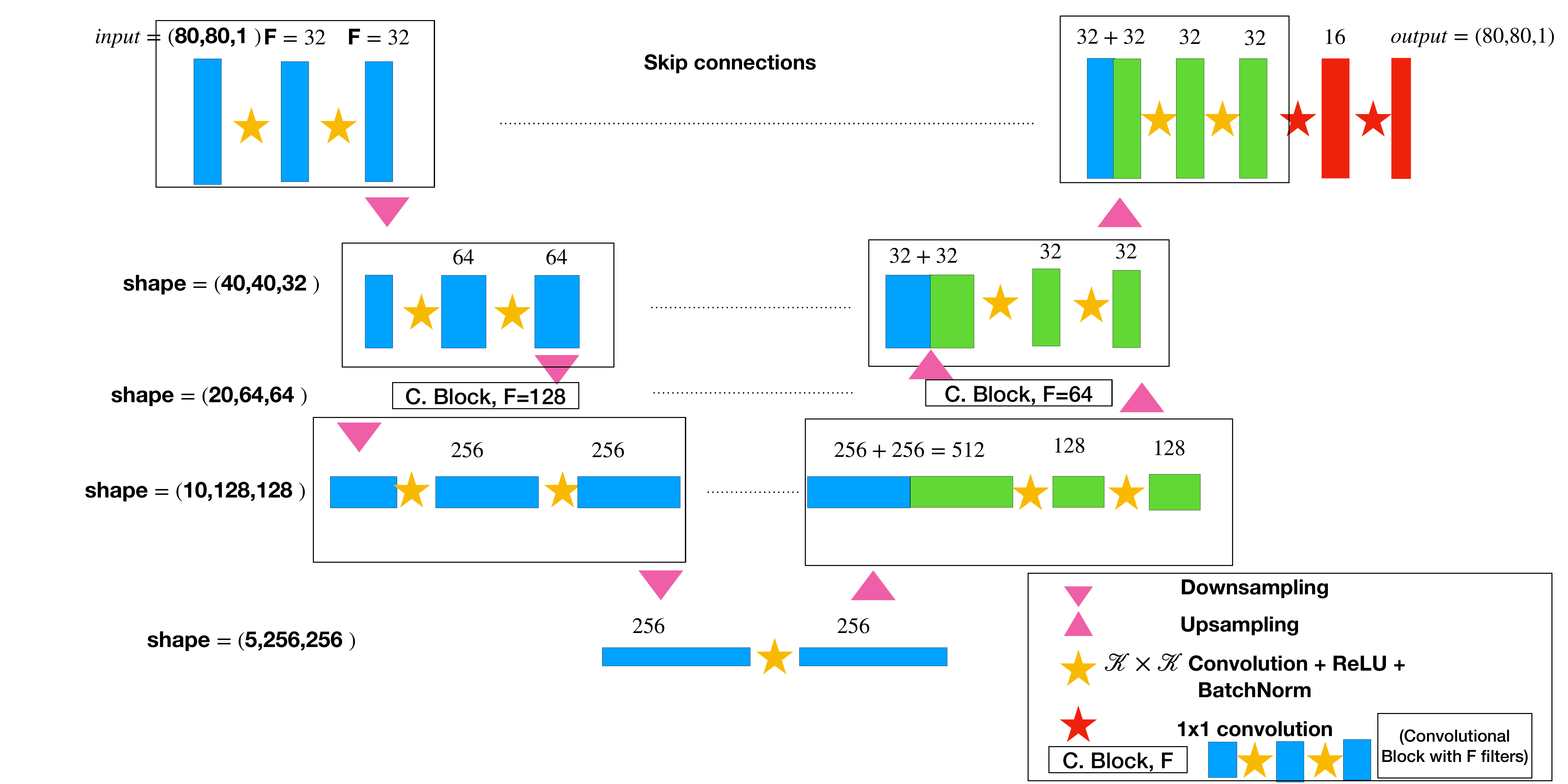}
\caption{U-Net architecture for input dimensions = (80 pixels, 80 pixels, 1 channel) and $\text{F}=32$. The input image (star, SZ or X-ray) is down-sampled 4 times, once per layer through a set of convolution of kernel size $\mathcal{K}\times \mathcal{K}$. At the very bottom, the down-sampled representation is up-sampled using a similar convolutional block to generate the output mass map. Skip connections are used to ensure that the information is not totally lost during the down-sampling operations. We remind that dropout is applied to all convolutional layers only in the decoder.}\label{fig:UNET}
\end{figure*}

The chosen model and its architecture used for the generator of mass density maps is based on convolutional neural networks following a U-Net architecture. This model was originally developed for image segmentation in the field of biological imaging \citep{UNET}. In astronomy, some applications followed the original purpose of segmentation such as removing radio frequency interference \citep{Akeret2017}, but also the U-Net has been successfully considered for various and different tasks \citep{Milletari2016,Aragon2019,Berger2019,Hausen2020,Lauritsen2021,Hong2021}. 
The U-Net is characterised by a set of two paths: the down-sampling path (or encoder) and the up-sampling path (or decoder). For our application, the considered U-Net is described below and represented in \autoref{fig:UNET}.
\begin{itemize}
    \item The \textbf{down-sampling path} consists of a succession of convolutional blocks, each of these applies two $\mathcal{K}\times\mathcal{K}$ convolutions with $F$ filters ($C_{\mathcal{K}\times{\mathcal{K}}}^{F}$), being $\mathcal{K}$ the size of the receptive field or the kernel size. After the convolutions, a rectified linear unit operation is applied \citep[ReLU; ][]{relu} to ensure nonlinearity, and a batch normalisation operation \citep[BatchNorm; ][]{batchnorm} is applied after each of the convolutions to improve the stability and performance of the network. Later, a 2x2 max-pooling \citep[e.g., ][]{maxpooling} operation down-samples the tensor shape so that this is reduced by a factor of $1/2$. The subsequent convolutional blocks are built with the same architecture, maintaining the kernel size of the convolutions but increasing the number of filters by a factor of 2, i.e., for the $n$ layer, the filters are $F\times 2^{n-1}$, where $n=1,..., N$ is the layer and $N$ is the maximum number of layers. After down-sampling $N$ times, a final convolution $C_{\mathcal{K}\times{\mathcal{K}}}^{F\times{2^{N-1}}}$ is applied. Therefore, the down-sampling path can be written as a series of convolutions as follows.
    \begin{multline}
        \text{Encoder}=C_{\mathcal{K}\times{\mathcal{K}}}^{F}C_{\mathcal{K}\times{\mathcal{K}}}^{F}\xrightarrow[\text{sampling}]{\text{down}}
       C_{\mathcal{K}\times{\mathcal{K}}}^{F\times 2}C_{\mathcal{K}\times{\mathcal{K}}}^{F\times 2}\xrightarrow[\text{sampling}]{\text{down}}\text{...}\\
       \text{...}\xrightarrow[\text{sampling}]{\text{down}}C_{\mathcal{K}\times{\mathcal{K}}}^{F\times{2^{N-1}}}C_{\mathcal{K}\times{\mathcal{K}}}^{F\times{2^{N-1}}}\xrightarrow[\text{sampling}]{\text{down}}C_{\mathcal{K}\times{\mathcal{K}}}^{F\times{2^{N-1}}}
    \end{multline}
    
    \item The \textbf{up-sampling path} consists of a succession of similar convolutional blocks to infer the output mass density map from the latent (central layer) representation given by the down-sampling path. From this encoded reduced representation, which is a tensor whose shape has been decreased by a factor of $1/2^{N}$ times and has $F\times 2^{N-1}$ channels, up-sampling operations are applied until the shape of the output mass density map is recovered. This up-sampling operation consists of repeating the nearest points to increase the shape of the data by a factor of 2. Then, skip connections are used for the same row layers, and thus, information is not bottlenecked in the latent representation. At the final decoder step, $1\times1$ convolutions are applied to recover the filter dimensions of the output mass density $\text{F}=1$. For preventing overfitting, random ``dropout'' \citep{dropout} is considered only in the convolutional blocks of the decoder.
    
    \begin{multline}
        \text{Decoder}=\xrightarrow[\text{sampling}]{\text{up}}C_{\mathcal{K}\times{\mathcal{K}}}^{F\times 2^{N-2}}C_{\mathcal{K}\times{\mathcal{K}}}^{F\times 2^{N-2}}
        \xrightarrow[\text{sampling}]{\text{up}}
       C_{\mathcal{K}\times{\mathcal{K}}}^{F\times 2^{N-3}}C_{\mathcal{K}\times{\mathcal{K}}}^{F\times 2^{N-3}}\\
       \xrightarrow[\text{sampling}]{\text{up}}\text{...}
      \xrightarrow[\text{sampling}]{\text{up}}C_{\mathcal{K}\times{\mathcal{K}}}^{F}C_{\mathcal{K}\times{\mathcal{K}}}^{F}C_{\text{1}\times{\text{1}}}^{F/2}C_{\text{1}\times{\text{1}}}^{1}
    \end{multline}

\end{itemize}

\subsubsection{Multiview approaches}
\begin{figure}
\includegraphics[width=\columnwidth]{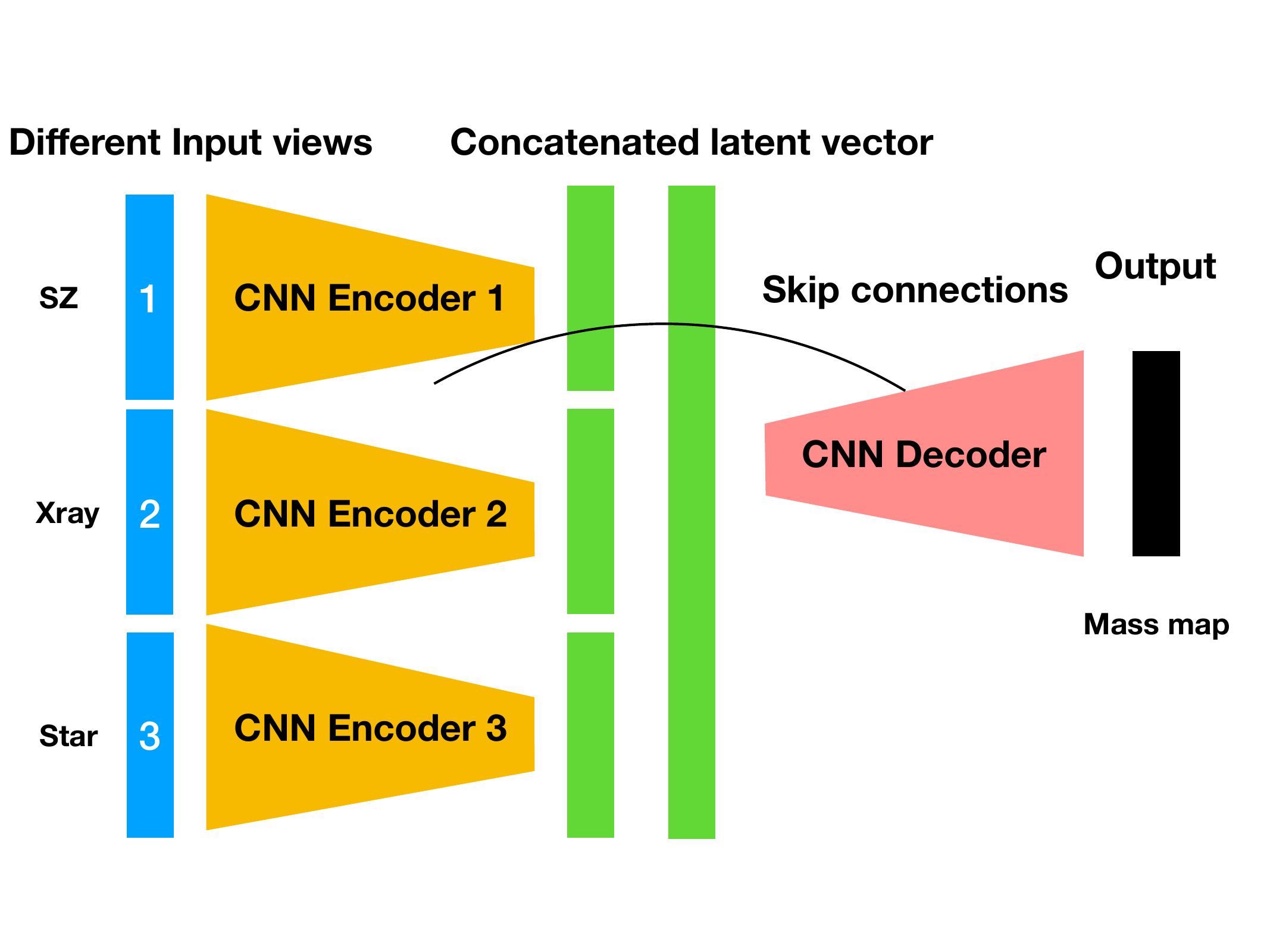}
\caption{Multiview 3 approach. 3 different encoders are considered, with one down-sampling path per each input view. Then this information is concatenated in the internal layer. From this concatenated latent vector, one decoder is in charge of the inference of the output mass density map.}\label{fig:multiview}
\end{figure}

This model has the advantage that it can be easily generalised to extract information simultaneously from multiple input views. To do that two approaches are studied:
\begin{enumerate}
    \item The three different views are combined in a single input tensor whose shape is increased as if it were an RGB image. This approach is labelled as ``multi-1''; and
    \item Three different encoders are used, one encoder for each input view. Subsequently, the three latent vectors are concatenated in the internal latent space. One decoder, the generator, is devoted to creating mass density maps from the information of this concatenated latent space. This approach is labelled as ``multi-3'' (see \autoref{fig:multiview}). 
\end{enumerate}
A summary of all the models used in this work can be found in \autoref{tab:models}.

\begin{table}
\begin{tabular}{@{}ccc@{}}
\toprule
Model name & input maps         & number of encoders \\ \midrule
star       & star               & 1                  \\
SZ         & SZ                 & 1                  \\
X-ray      & X-ray              & 1                  \\
multi-1    & star, SZ and X-ray & 1                  \\
multi-3    & star, SZ and X-ray & 3                  \\ \bottomrule
\end{tabular}\caption{Models considered in this work. Each is a variation of the U-Net architecture presented in \autoref{fig:UNET} to account for different inputs.}\label{tab:models}
\end{table}

\subsection{Training and validation}

 According to the description given in the previous section, we consider the following hyperparameters for our model: the number of channels or filters $(F)$, the size of the kernel in the convolutions (K), the number of layers in both the encoder and decoder architectures $(N)$, and the fraction of neurons that are randomly omitted, i.e. the  ``dropout''. The considered possible values of these hyperparameters are shown in \autoref{tab:hyperparameters} and the optimal set of hyperparameters is found by {\sc{Tree of Parzen Estimators }} algorithm \citep[TPE; ][]{TPE}. This algorithm is implemented and freely available at {\sc Hyperopt\footnote{\url{http://hyperopt.github.io/hyperopt/}}: Distributed Asynchronous Hyper-parameter Optimisation} \citep{hyperopt}. The targeted loss function is the mean absolute error $L_1$ between the predicted pixel values and the ground-truth pixels. 

 Furthermore, the data is randomly split into 3 subsets: the training dataset composed of $80\%$ of the sample, the validation dataset with $10\%$ of data and the remaining $10\%$ belongs to the test dataset. The training dataset is used for tuning the U-Net parameters, the {\sc{Tree of Parzen Estimators }} algorithm is fed with validation data, and the test set is only used for displaying the final results.

 We have taken 100 evaluations when performing the {\sc Hyperopt} optimisation, which means that the U-Net needs to be fitted 100 times for each of the 5 different U-Nets. The models are trained using the {\sc Adam} optimiser \citep{kingma2014:Adam} with a learning rate of $10^{-4}$, which is reduced to $10^{-5}$ after 50 epochs. The training stops if the validation loss does not improve after 5 epochs, a process commonly known as ``early stopping'', or if a model has trained 100 epochs. The hardware used for training is an NVIDIA A100 GPU, which translates to 1 epoch and takes about 2 minutes to complete. Before fitting the models, the maps are normalised following the common standard normalisation, see e.g., \cite{deAndres2022Planck} for further details.

\begin{table}
\centering
\begin{tabular}{@{}ll@{}}
\toprule
hyperparameter     & values       \\ \midrule
Filters $(F)$          & 8, 9, ..., 32    \\
Kernel (K)           & 1, 2, ..., 6      \\
Number of layers $(N)$ & 1, ..., 4      \\
dropout            & {[}0, 0.5{]} \\ \bottomrule
\end{tabular}
\caption{The considered possible values of hyperparameters of our U-Net model.}
\label{tab:hyperparameters}
\end{table}

\section{Results}\label{sec-4}
\begin{figure*}
\includegraphics[width=\textwidth]{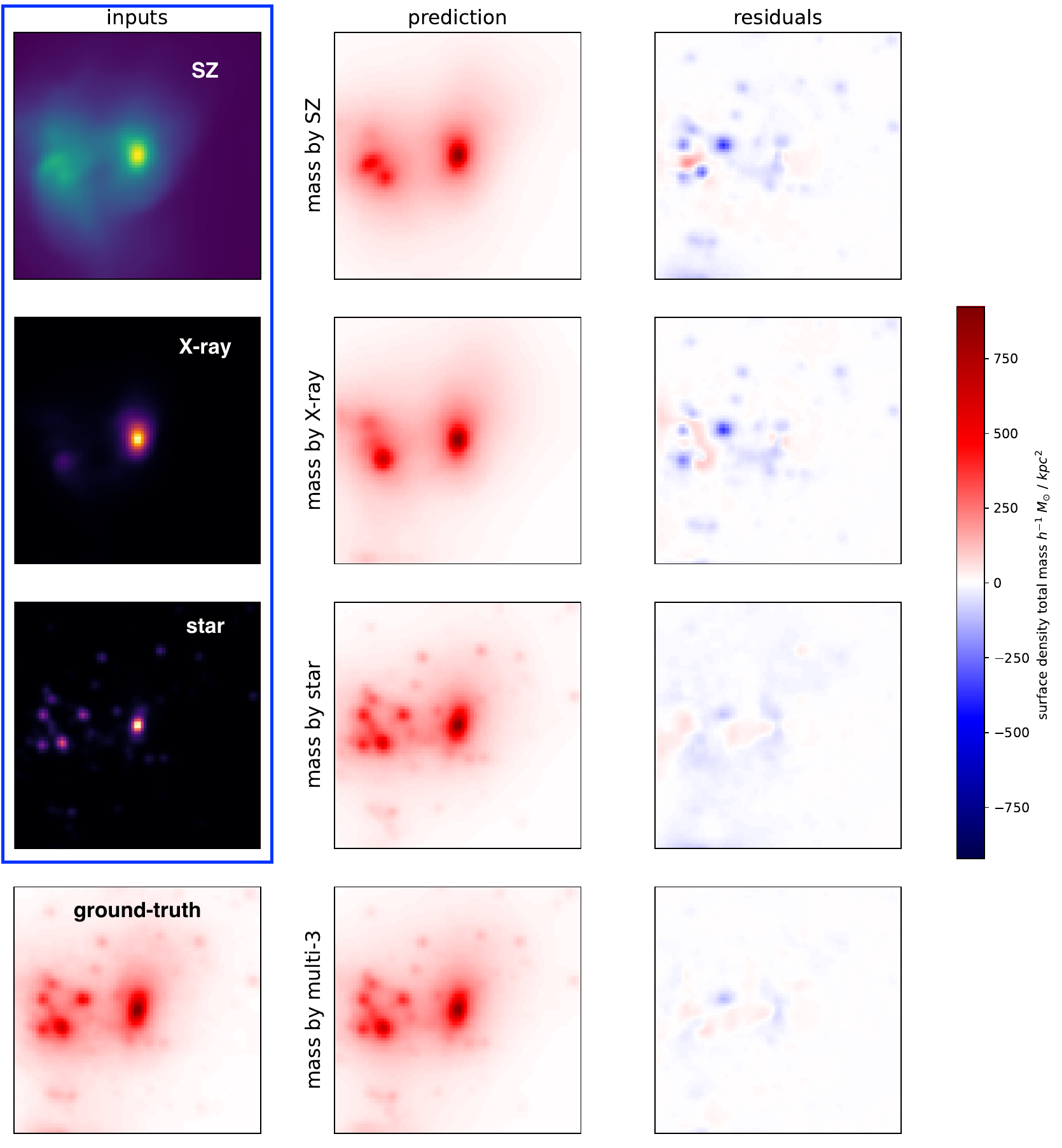}
\caption{The first column on the left corresponds to the input maps (SZ, X-ray and stars) and the ground-truth mass density map. In the second column from top to bottom, we show the mass map predictions of our U-Net when training with SZ, X-ray, star or multiview (multi-3). The residuals are defined as the difference between the prediction and the ground-truth maps. The size of all maps is $2\times R_{200}$.}
\label{fig:maps_output}
\end{figure*}

In this section, our aim is to assess the quality of predicted density maps by our model. To do that, we compare the ground-truth density maps with their corresponding predicted maps in the test set with simple visualisation as the first metric for assessing the quality of the predictions. Firstly, we show one example of our predicted maps in \autoref{fig:maps_output} for our different U-Nets accounting for different input views. The first column on the left shows the three input views corresponding to the same galaxy cluster and the ground-truth density map is located at the top left. The second column shows the predictions for several input bands: SZ, X-ray, star and multiview. The last column corresponds to the residuals, i.e., the difference between the prediction and the ground truth. 

As a general result, we observe that the predicted mass density maps from SZ and X-ray inputs are smoother and do not contain most of the substructures that can be appreciated in the ground-truth map. This is clear in the last column where the residuals mostly contain all the missing substructures and they are underestimated as shown in the figure in blue colour. Conversely, predictions from stars and from the multiview models contain most of the substructures. Their residuals are generally closer to zero than the others. We only show here the multi-3 approach, given the similarity among multi-1 and multi-3 predictions in a human eye test.

This visualisation of density-map residuals in \autoref{fig:maps_output} apparently is not sufficient to numerical quantify the similarity between predict and true maps. Therefore, we are compelled to utilise a set of additional metrics to measure the discrepancies between the two. The considered metrics are the pixel-wise statistics, cylindrical radial mass profiles, power spectrum and maximum mean discrepancy which are studied in the following subsections.

\subsection{Pixel-wise statistics}

\begin{figure*}
\includegraphics[width=1.0\textwidth]{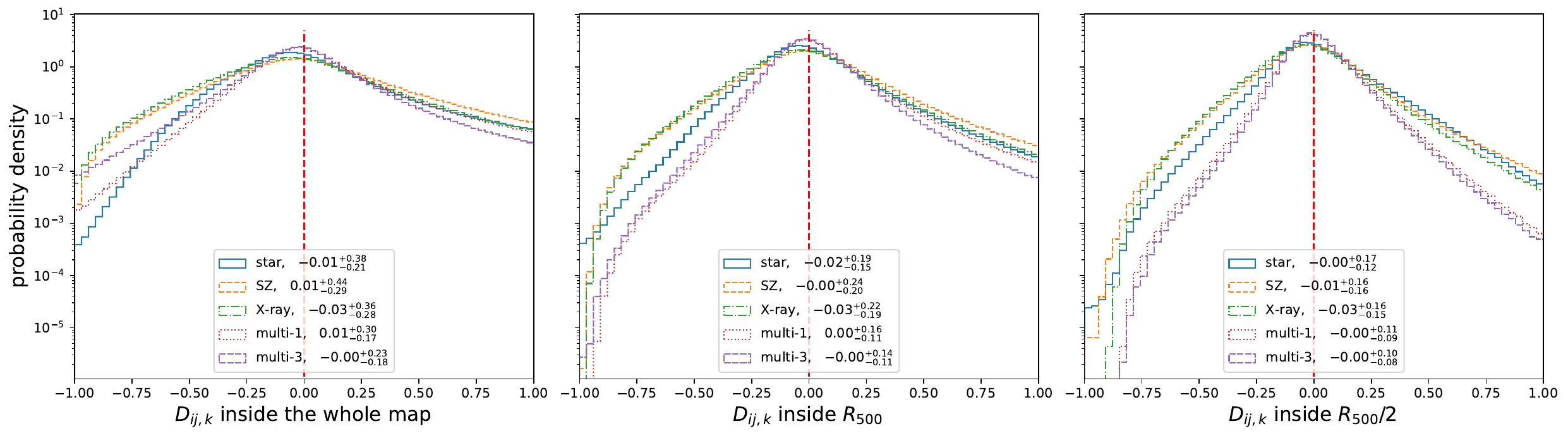}
\caption{Pixel-wise relative difference $D_{ij,k}$ (see Eq.(\ref{eq:pixel_diff_Dijk})) between the predicted mass density maps $\hat{I}_{ij,k}$ and the ground-truth mass density maps $I_{ij,k}$. Different lines represent the use of different input views to predict the mass density map: star, SZ, X-ray, multi-1 and multi-3. The legend also shows the median value of the distributions with the $16^{\text{th}}$ and $84^{\text{th}}$ percentiles as $\text{median}^{+|84^{\text{th}}-\text{median}|}_{-|\text{median}-16^{\text{th}}|}$. From left to right, $D_{ij,k}$ is computed for pixels inside various circular apertures: the whole map, inside $R_{500}$ and inside $R_{500}/2$.}\label{fig:pixel_stats}
\end{figure*}

One interesting metric is considering the pixel-value differences between ground truth and predicted maps. We calculate the relative difference $D_{ij}$ between predicted maps $\hat{I}_{ij}$ and true maps $I_{ij}$ as
\begin{equation}\label{eq:pixel_diff_Dijk}
    D_{ij,k} = \frac{\hat{I}_{ij,k}-I_{ij,k}}{I_{ij,k}}\text{ .}
\end{equation}

Therefore, the tensor $D_{ij,k}$ computes the pixel-wise similarity between true images and predicted images. The index $k$ here runs over maps, i.e., all the maps in the test set. Subsequently, the tensor $D_{ij,k}$ is flattening to a vector of dimensions $i\times j \times k$ and its histogram as a probability density is represented in \autoref{fig:pixel_stats}.  We drop the values of the tensor $D_{ij,k}$ where the ground-truth signal $I_{ij,k}$ equals 0 unless $\hat{I}_{ij,k}-I_{ij,k}=0$.

The particular values of the median, $16^{\text{th}}$ and $84^{\text{th}}$ percentiles of the distributions are also displayed in the legend of \autoref{fig:pixel_stats} as: 
\begin{equation}\label{eq:scatter}
    \text{value}^{+\text{error}}_{-\text{error}} = \text{median}^{+|84^{\text{th}}-\text{median}|}_{-|\text{median}-16^{\text{th}}|}. 
\end{equation}

In this part, we also study the performance of each input map to infer the predicted mass map at different apertures around the cluster. In the left panel, pixel values inside the whole map are considered, in the middle only the values $D_{ij,k}$ inside $R_{500}$ are displayed and in the right panel only values inside $R_{500}/2$. Firstly, for the three panels in \autoref{fig:pixel_stats}, the same trend is observed: the multiview-3 is the most accurate, followed very closely by the multiview-1, using star density maps as input corresponds to the third most accurate model while considering X-ray and SZ as inputs results in a higher relative error, which is consistent to the expectation from \autoref{fig:maps_output}. Furthermore, this can be appreciated in the legend of \autoref{fig:pixel_stats} that all the models are slightly biased mostly towards negative values, besides the multiview-3 with $\text{median}\simeq 0.00$ in all three panels. While the model trained with only X-ray images yields the worst with $\text{median}=-0.03$ for all three panels. Moreover, the scatter is smaller for the two multiview approaches, i.e., $+0.23,-0.18$ with only slightly larger in multiview-1 than in multiview-3. In contrast, that scatter from the model trained with only SZ images is the largest, $+0.44,-0.29$, followed by the X-ray model, $+0.36,-0.28$, which is also closer to the values from the star model, $+0.38,-0.21$. As it is depicted in \autoref{fig:maps_output}, models whose input is either SZ or X-ray fail at predicting small-scale substructures and the signal vanishes in the outskirts of the clusters. Therefore, all those not predicted substructures contribute to negative values in the $D_{ij,k}$ distribution. Subsequently, these substructures in the density mass maps are, allegedly, wrongly estimated. This will be quantified in the subsequent sections.

By comparing with the middle and right panels of \autoref{fig:pixel_stats}, we aim to test whether the asymmetry in the distributions of the relative difference tensor $D_{ij,k}$ is mainly affected by pixels outside the central region of the maps and whether the model behaves worse in the cluster centre or not. As depicted in that figure, the distributions progressively become less spread and more symmetrical when considering only pixels inside a smaller aperture, $R_{500}$ and $R_{500}/2$. This means that the central regions, which we care the most, actually have the best result. Nevertheless, the multiview models provide the most accurate result in all cases, albeit a slight improvement with multiview-3 than multiview-1.

\subsection{Cylindrical radial profiles}
\label{sec:profiles}

\begin{figure*}
\includegraphics[width=0.9\textwidth]{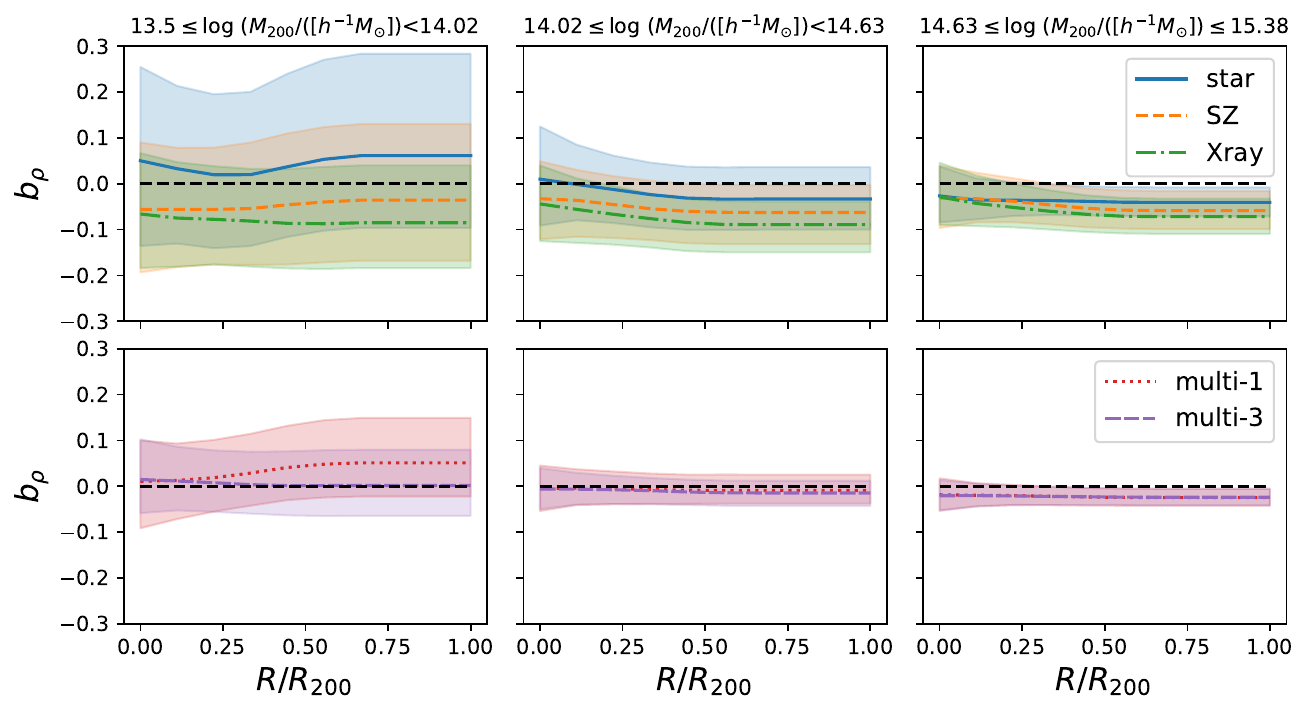}
\caption{Radial profiles of the mass bias, see Eq.(\ref{eq:bias}). From left to right, we show the bias corresponding to different mass intervals as indicated in \autoref{fig:massstats}: Interval 1 corresponds to $13.5\leq\log M_{200}/\hMsun<14.02$, the interval 2 range is $14.02\leq\log M_{200}/\hMsun<14.63$ and the interval 3 is $14.63\leq\log M_{200}/\hMsun\leq15.38$. The top panels show the single-input view models (star, SZ, X-ray) and the bottom panels the multiview-1 and -3 models. The lines represent the median values per bin and the shaded regions cover the $16^{\text{th}}$ and $84^{\text{th}}$ percentiles. Furthermore, the particular enclosure mass bias at $R_{200}$ is presented in \autoref{tab:mass_profiles}.}\label{fig:mass_profiles}. 
\end{figure*}

One can assess the quality of the predictions by comparing the mass radial profiles of the predicted and the ground-truth mass density maps. The mass profiles $M_{\rho}$ are defined by integrating the mass density, or pixel values, of the maps in different circular regions $\Omega_{\rho}$ with radius $\rho$:
\begin{equation}\label{eq:mass_profiles}
    M_{\rho}= \sum_{i,j \in \Omega_{\rho}} I_{ij}\Sigma \text{ .} 
\end{equation}
Note that all circular regions include pixels from the very centre of the maps, i.e., $\text{radius}\in [0,\rho]$. Here, $I_{ij}$ are the pixel values and $\Sigma$ is the pixel size -- $\Sigma = \left(R_{200}/40\right)^{2}$ --. Note that Eq.(\ref{eq:mass_profiles}) provides the total 3D mass of a galaxy cluster integrated over a cylindrical volume of $2\times R_{200}\times \Omega_{\rho}$. We then have computed $M_{\rho}$ for ten equally-spaced circular regions of radii $\rho=R_{200}/10,\text{ }R_{200}/9,\text{ }...,\text{ }R_{200}$. Nevertheless, the difference between the ground-truth profile $M_{\rho}$ and the predicted profile $\hat{M}_{\rho}$ can be estimated as the mass bias $b_{\rho}$ as it is computed also in \cite{deAndres2022Planck} and \cite{Ferragamo2023}:
\begin{equation}\label{eq:bias}
    b_{\rho} = \frac{\hat{M}_{\rho}-M_{\rho}}{M_{\rho}}\text{ .}
\end{equation}

In \autoref{fig:mass_profiles}, we show the bias $b_{\rho}$ as a function of the normalised radius $R/R_{200}$. From left to right, we show the bias that corresponds to three different mass intervals, which divide the dataset in a manner that each interval contains roughly the same number of galaxy clusters, see \autoref{fig:massstats}. Therefore, Interval 1 corresponds to $13.5\leq\log M_{200}/\hMsun<14.02$, the interval 2 range is $14.02\leq\log M_{200}/\hMsun<14.63$ and the interval 3 is $14.63\leq\log M_{200}/\hMsun\leq15.38$. Moreover, in the top panels, we show the 3 single-view approaches (star, SZ, X-ray) and the 2 multiview models are shown in the bottom panels. From \autoref{fig:mass_profiles}, we see that all models have negative biases, apart from the low mass interval in which the biases for SZ and Xray models generally are negative, but positive for star and multi-1 models. For the single view result in the top panels of \autoref{fig:mass_profiles}, negative biases for the results from SZ and X-ray inputs are possibly caused by the fact that they can not provide any information of substructures, especially relatively low mass ones. The biases for the stellar inputs decrease from positive at the low halo mass bin to negative at the high mass bin. We suspect that is because the contribution of substructures is more significant in halos with higher mass. Since we train the model regardless of the sample's halo mass, there is also a possibility that ML has to balance the predictions due to different contributions of the sample's halo mass. Nevertheless, it is surprising to see the improvements from multiview inputs shown at the bottom panels of \autoref{fig:mass_profiles}: the median bias is reduced by roughly a factor of $\sim 1/3$, while the scatter significantly shirked by almost $\sim 1/2$ in all three mass bins. For the reader's interest, the particular values for the bias at $R_{200}$ for different models and mass intervals are shown in \autoref{tab:mass_profiles}. 

The data in \autoref{fig:mass_profiles} and \autoref{tab:mass_profiles} show a general trend that the scatter of the bias decreases with mass regardless of the input images. As can be seen in the Table, for interval 1 considering the multiview-3 model, the results are $b_{200}=0.001^{+0.078}_{0.066}$ and for interval 3 is $0.024^{+0.019}_{-0.017}$, that is, the width is reduced by a factor of $4$. This translates into the effect that the predictions for massive clusters are more precise than those predictions for less massive clusters with smaller scatter. This problem could be caused by different factors:
\begin{itemize}
    \item The loss function utilised during training, which is the mean absolute error $L_{1}$, could be inducing the weights of the neural network to be updated to fit only massive clusters. This can be understood by noticing that the gradients of the loss function could be higher for massive clusters. According to this hypothesis, the presence of high-mass clusters in our dataset hinders the predictions at low masses. We claim here that this is false as demonstrated in the Appendix \autoref{sec:appendix}, in which the deep learning model is trained with a different configuration where there is only data corresponding to low masses in the training set.
    \item ICM and stars follow gravity at higher masses, but at lower masses astrophysical effects are relevant and ICM and star components differ between \gadgetx\ and \simba\ simulations, increasing the irreducible scatter in the observable-mass relation. Since the training uses both simulation halos together, this larger difference at lower halo mass introduced such a large scatter. Furthermore, even with a single simulation, the scatter in halo baryon fractions increases as halo mass decreases which is caused by their intrinsic evolution history \citep[see][for example]{Cui2021}. This is the most relevant source that induces a mass-dependent scatter, as it is discussed in further detail in the Appendix \ref{sec:appendix}.
\end{itemize}


\bgroup
\def\arraystretch{2}
\begin{table}
\begin{tabular}{@{}cccc@{}}
\toprule
model       & Interval 1                & Interval 2                   & Interval 3                  \\ \midrule
star        & $-0.061^{+0.222}_{-0.158}$  & $-0.033^{+0.069}_{-0.066}$ & $-0.041^{+0.033}_{-0.030}$ \\
SZ          & $-0.036^{+0.166}_{-0.132}$  & $-0.063^{+0.060}_{-0.069}$ & $-0.059^{+0.042}_{-0.040}$ \\
X-ray        & $0.085^{+0.125}_{-0.098}$   & $-0.089^{+0.050}_{-0.060}$ & $-0.071^{+0.039}_{-0.038}$ \\
multi-1 & $0.051^{+0.098}_{-0.073}$   & $-0.009^{+0.034}_{0.028}$  & $-0.025^{+0.019}_{-0.018}$ \\
multi-3 & $0.001^{+0.078}_{-0.066}$   & $-0.015^{+0.028}_{-0.026}$ & $-0.024^{+0.019}_{-0.017}$  \\ \bottomrule
\end{tabular}
\caption{Values for the bias $b_{200}$ defined in of Eq.(\ref{eq:bias}) at $R_{200}$ for our different models. These values are also displayed in \autoref{fig:mass_profiles} and the intervals 1, 2 and 3 are defined to separate the dataset in roughly equal number of clusters, as shown in \autoref{fig:massstats}. Consequently, Interval 1 corresponds to $13.5\leq\log M_{200}/\hMsun<14.02$, the interval 2 range is  $14.02\leq\log M_{200}/\hMsun<14.63$ and the interval 3 is $14.63\leq\log M_{200}/\hMsun\leq15.38$.}
\label{tab:mass_profiles}
\end{table}
\egroup

\subsection{Power spectrum}
Quantifying the information contained in our maps at different spatial frequencies is of great interest because the empirical evidence suggests that substructures tend to be underestimated as observed in \autoref{fig:maps_output}, especially for the X-ray and SZ input images. This can be clearly viewed by investigating the power spectrum of the density mass map at small scales. For 2D images ($I_{nm}$) of pixel size $N\times M$, the Fourier Transform $\mathcal{F}_{k_nk_m}$ can be defined as:

\begin{equation}
    \mathcal{F}_{k_xk_y} = \sum_{n}^{N}\sum_{m}^{M} I_{ab} \exp \left[
    -i2\pi\left(\frac{k_x n}{N}+\frac{k_y m}{M}\right) 
    \right]\text{ ,}
\end{equation}
where $i$ stands here for the imaginary unit. Note that for our images of $80\times 80$ pixels $N=M=80$. Moreover, the power spectrum is computed as:
\begin{equation}
    P_{k} = |\mathcal{F}_{k}|^{2} \left(\frac{2R_{200}}{80^{2}}\right)^{2}
\end{equation}
where $k=\sqrt{k_x^{2}+k_{y}^{2}}$. Note that the possible values of $k$ in pixels go from 1 to the one corresponding to the Nyquist frequency, i.e., $\sqrt{2}\times N/2\simeq 56$ pixels, or in physical units with a factor of $\pi/R_{200}$. From here we defined the spatial length $\lambda$ as
\begin{equation}
    \lambda = 2\pi/k \text{.}
\end{equation}

\begin{figure*}
\includegraphics[width=0.9\textwidth]{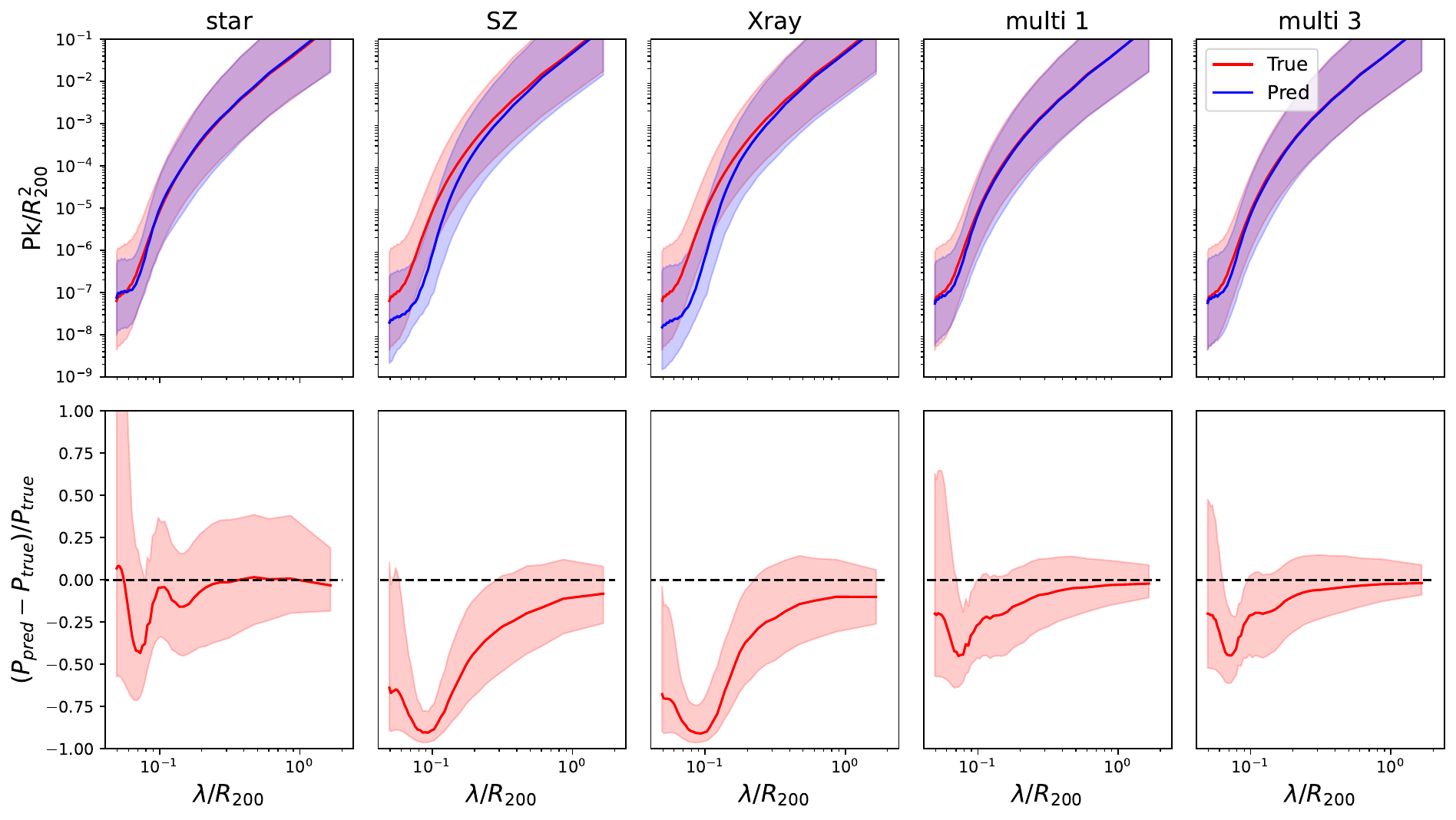}
\caption{\textbf{Top:} Power spectrum corresponding to our ground-truth (red) and the predicted (blue) mass density maps as a function of the spatial length $\lambda= 2\pi/k$ for our different inputs: star, SZ, X-ray, multi-1 and multi-3, are displayed in different columns. \textbf{Bottom:} We show the relative difference $(P_{\text{pred}}-P_{\text{true}})/P_{\text{true}}$ of the predicted power spectrum $P_{\text{pred}}$ and the ground-truth power spectrum $P_{\text{true}}$ of our mass density maps. The dashed black line depicts the perfect prediction where the difference is zero. The solid lines correspond to the median values while the shaded regions represent the $16^{\text{th}}$ and $84^{\text{th}}$ percentiles. }\label{fig:power}
\end{figure*}

Therefore, we further calculate the power spectra of the ground-truth maps and of the predicted maps for comparison, using the {\sc Pylians} library \citep[{\textbf{Py}thon \textbf{li}braries for the \textbf{a}nalysis of \textbf{n}umerical \textbf{s}imulations, }][]{Pylians},  and highlight the difference in \autoref{fig:power}. In this figure, the power spectra of the predicted maps and the ground-truth maps are shown in the top panels as a function of the spatial length which is normalised to $R_{200}$. In the lower panels, we show the relative difference of the spectrum defined as $(P_{\text{pred}}-P_{\text{true}})/P_{\text{true}}$. In this figure, several things can be noted. Firstly, the power spectrum of the predicted mass density maps from star maps is consistent within the errors to the ground truth, where the median power spectrum starts deviating significantly at scales $0.1\times R_{200}$. For ICM tracers, the mass density maps predicted from SZ and X-ray observables indicate that their power spectrum differs on average from the ground-truth one (lower by about 25 per cent at $\lambda \sim 0.5 R_{200}$), being only similar at high spatial wavelengths. In addition, in \autoref{fig:power} the advantage of considering multiview models is appreciated, as shown in the fourth and fifth columns in the figure, where the multiview models effectively combine the information available from the different inputs. This means that the median power spectrum is similar to the power spectrum of the predicted mass density maps from stars, but with the clear advantage that the scatter is much smaller. To this end, SZ and X-ray views do not, as expected, improve the predictions of small structures but rather allow the model, when combined with data from stars, to better calibrate the overall signal, reducing the scatter.

\subsection{Maximum Mean Discrepancy}

\begin{figure}
\includegraphics[width=\columnwidth]{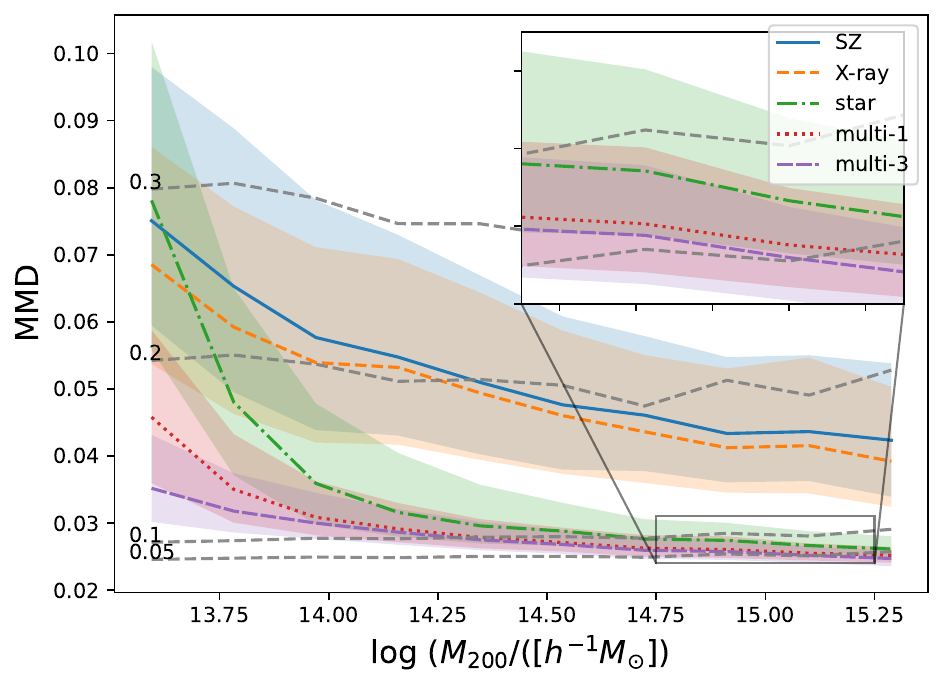}
\caption{Maximum Mean Discrepancy, see Eq.(\ref{eq:MMD}), between predicted mass density maps and the ground-truth mass density maps as a function of the cluster mass $M_{200}$. Different colours represent the predictions when the model is trained with only SZ data (blue), X-ray data (orange), stellar data (green), multi-1 (red) and multi-2 (purple). Horizontal grey dashed lines represent the calibration of the MMD values, Eq.(\ref{eq:callibrationMMD}), and numbers written in black colour on the left of these lines correspond to the noise intensity $\alpha$ used for the calibration of the MMD values. The inset highlights the results at higher halo mass where the multiview results are compatible with $\alpha \sim 0.05$. }\label{fig:MMDs}
\end{figure}

Given two distributions, maximum mean discrepancy (MMD) is a test that assesses whether the two images are the same \citep{MMD}. Although MMD can be used for training generative adversarial neural networks \citep{trainMMD}, it is used here for simply assessing the quality of the predicted mass maps. 

The MMD can be defined by choosing a kernel function $k$ and a pair of random variables of inputs $X$ and outputs $Y$, so that one can compute the MMD as 
\begin{multline}\label{eq:MMD}
    \text{MMD}^{2}(X,Y) = \frac{1}{m(m-1)}\sum_{i}\sum_{j\neq i} \mathcal{K} (x_{i},x_{j})-2\frac{1}{m^{2}}\sum_{i}\sum_{j} \mathcal{K}(x_i,y_j)\\
    + \frac{1}{m(m-1)}\sum_{i}\sum_{j\neq i } \mathcal{K}(y_i,y_j)\text{.}
\end{multline}
Here $x_{i}$'s are the ground-truth data points and $y_{i}$'s are the predictions of our U-Net model. For our case, the kernel is Gaussian, and therefore the estimation of $\mathcal{K}(x,y)$ can be written as
\begin{equation}\label{eq:GaussianKernell}
    \mathcal{K}(x,y)=\exp \left( \frac{-||x-y||}{2\sigma^2} \right).
\end{equation}
where $\sigma$ corresponds to the band-with range. Typically, one chooses a range of values of $\sigma$ to evaluate the MMD. In our case, we have computed the MMD as the maximum value estimated by Eq. (\ref{eq:MMD}) taking into consideration the following values of $\sigma = [0.01, 0.1, 0.25, 0.5, 0.75, 1.0, 2.5, 5.0, 7.5, 10.0]$, defined in Eq. (\ref{eq:GaussianKernell}).

Though the MMD might be very useful, it has little physical meaning and needs to be calibrated. To overcome this issue, we have computed the MMD between the ground-truth mass density map ${I}_{ij}$ and a perturbed map $P_{ij}$ that is created by adding a random Gaussian noise such that 
\begin{equation}\label{eq:callibrationMMD}
    P_{ij} = I_{ij}+\mathcal{N}(0,\max (I_{ij})\cdot \alpha)\text{ .}
\end{equation}
Here $\alpha$ corresponds to the parameter that quantifies the standard deviation of the noise map. In \autoref{fig:MMDs}, we have used $\alpha=0.05,0.1,0.2,0.3$ as shown in horizontal grey dashed lines, which represent the calibration of the MMD. This means that the standard deviation of the noise map is at most 30\% of the peak value. 

Subsequently, we have calculated the MMD between the ground-truth maps $I$ and the predicted maps $\hat{I}$. In \autoref{fig:MMDs}, we show the median values of the MMDs as a function of the mass of the cluster that corresponds to our different models. As a general result, the MMD metric is in agreement with the other previous investigations in the sense that the multi-3 model is the closest to the ground truth. We also find a mass dependence of the MMDs values with respect to the total mass of the galaxy clusters, which is coherent with the results in \autoref{sec:profiles} and it is explained in detail in the Appendix \ref{sec:appendix}. Moreover, the noise calibrations, which are represented in dashed grey lines, show that for small groups of clusters, the discrepancy is equivalent to having less than $30\%$ of noise in the maps. However, in the massive end of our dataset around $10^{15}\hMsun$, MMDs are below $10\%$ for all the models (SZ, X-ray, star, multi-1 and multi-3) and very close to $5\%$ in case of the multiview models. This result suggests the use of multiview data to boost the statistical agreement of the pixel's distribution in the generated total mass density maps.

\section{Conclusions}\label{sec-5}

Deep Learning models have been used for generating images \citep[see][for example for the application in astronomy]{Rothschild2022}. Building on the success in the development of ML in this field, we propose, for the first time, to expand its application to predicting the total matter density distributions using multiview simulated observational data: star, Sunyaev-Zeldovich and bolometric X-ray. The deep learning architecture used on the project is based upon the U-Net, which was introduced in the context of biomedical imaging and it has been modified for accounting for multiview input as described in \autoref{sec:Model}. As the first step along this research line, we validated the applicability of this ML model using a dataset of simplified maps, described in \autoref{sec-2}, from {\theth{}} set of hydrodynamic simulations. This dataset is free from instrumental and observational effects so that noise, point sources and telescopes' impact are not considered. In this work, we tested different convolutional U-Net architectures for both single-input and multi-input models and applied different matrices to quantify the fidelity of the predicted matter density maps from this DL model. 
From the results, we can conclude that, although the ML output is much more complex -- 2D vs 1D or a single data point ($M_{200}$) -- compared to previous ML models, its outcomes, for example the 1D mass bias profile, are much better and very promising, especially for the multiview inputs. In detail, we can summarise our results in the following:
 \begin{itemize}
     \item The scatter in the estimation of the mass maps from multiview is reduced by a factor of $\simeq 1/2$. As depicted in \autoref{fig:pixel_stats}, the pixel statistics show that the scatter can be reduced from, e.g. $ \pm \sim16\%$ when inferring the mass from SZ and $ \pm \sim 10\%$ when using the combination of inputs maps using the multiview-3 model. Note that the scatter is defined as the deviation from the median value using the $16^{\text{th}}$ and $84^{\text{th}}$ percentiles as it is written in Eq.(\ref{eq:scatter}). This fact is also acknowledged when examining the mass profiles in \autoref{fig:mass_profiles} in which the scatter can be reduced from $\pm \sim 4\%$ to $\pm \sim 2\%$, see \autoref{tab:mass_profiles} for further details. 
     \item Each input view (see \autoref{fig:maps_output}) distinctly correlates with the output mass map and therefore, the capability of predicting different spatial frequencies is supreme for the multi-input models, as it is manifested in \autoref{fig:power}.
     \item By computing the MMD values defined in Eq.(\ref{eq:MMD}), we have examined the statistical similarity of the pixel distributions between the predicted maps and the true maps. The results suggest again the multiview models are foremost and they can be below $5\%$ noise level as shown in \autoref{fig:MMDs}.
     \item Overall, the multiview models provide the best predictions in all the matrices: very low bias with a small scatter, especially in the massive regime. Multi-3 shows a little improvement compared with multi-1 because using 3 different encoders tends to better capture the important features of each input image.
 \end{itemize}

 It is interesting to also asses what is the improvement in the mass reconstruction when combining only gas inputs, i.e., SZ and X-ray. For that purpose we have trained a model with two encoders and the results are in accordance with single-input models, but with no observed improvement in the quality of the predicted mass. Quantitatively, this can be analysed by computing the mass bias in the mass profiles which has roughly the same values as considering SZ alone. The performance of the double-input gas model seems to be not comparable with the benefits of including information from both tracers, e.g., star and gas together. The star model trained with star density maps is more precise than the corresponding double-input one with SZ and X-ray.   
    
 The results of this work hint that observations of galaxy clusters at different spectral bands will translate into a better estimation of the underlying mass distribution. We have limited ourselves, as a first work on the topic, to 3 input tracers, but our deep learning approach can be generalised to account for many available inputs at different wavelengths, e.g., the several luminosity bands of the SDSS survey. The accessible computational resources are the only limitation, such as GPU memory. 
 
 Different Deep Learning models could also be considered as an additional job. Though not shown in this paper, we have considered modifying the loss function in the framework of generative adversarial networks \citep{Goodfellow:GAN}, and training the Wasserstein GAN \citep{WGAN} with no improvement over using only the MAE loss function. Therefore, at the moment, the U-Net model presented in this work is, the most suitable for addressing the problem of the inference of the mass map from multiview simulated observations. Conversely, vision transformers \citep{ViT} might yield competitive results in comparison with convolutional networks.

 These techniques that are based on image-to-image translation algorithms \citep{Isola2016} can also be applied to painting baryons on to N-body simulations \citep{Chadayammuri2023}, which we will also consider them for generating baryon maps from DM-only simulations, increasing the number of maps already available in simulations.

 The results of this project represent the required proof-of-concept step towards the estimation of mass density maps from real observational data. This constitutes a challenge in the sense that our data considered here for training the models could not completely match the characteristics of real data due to several regards. Firstly, our models can be used in real data training on mock observational data where the instrumental impacts are considered in the framework of Simulation-Based Inference. As achieved in our previous work, observational mock data was created by introducing directly the instrumental effects on our simulated clean maps \citep{deAndres2022Planck}. Secondly, the physics implemented in our simulations could lead to biased results. To address this problem, deep learning models could be trained with data provided from simulations with various and plausible physical models \citep{CAMELS}.

\section*{Acknowledgements}

We would like to express our gratitude to the anonymous referee for the comments that greatly helped us improve the quality of the manuscript. D.d.A., W.C. and G.Y. would like to thank Ministerio de Ciencia e Innovación for financial support under project grant PID2021-122603NB-C21. WC is supported by the STFC AGP Grant ST/V000594/1 and the Atracci\'{o}n de Talento Contract no. 2020-T1/TIC-19882 granted by the Comunidad de Madrid in Spain. He also thanks the ERC: HORIZON-TMA-MSCA-SE for supporting the LACEGAL-III project with grant number 101086388 and the China Manned Space Project for its research grants.  MDP and AF acknowledge support from Sapienza Università di Roma thanks to Progetti di Ricerca Medi 2022, RM1221816758ED4E. The authors acknowledge The Red Española de Supercomputación for granting computing time for running the hydrodynamic simulations of The300 galaxy cluster project in the Marenostrum supercomputer at the Barcelona Supercomputing Center. 

\section*{Data Availability}

The results shown in this work use data from The {\sc Three Hundred} galaxy clusters sample. The data is freely available upon request following the guidelines of The Three Hundred collaboration, at  \url{https://www.the300-project.org}.



\bibliographystyle{mnras}
\bibliography{main} 




\appendix
\section{The differences between \simba{} and Gadget-X}
\label{sec:appendix}

\begin{figure}
\includegraphics[width=\columnwidth]{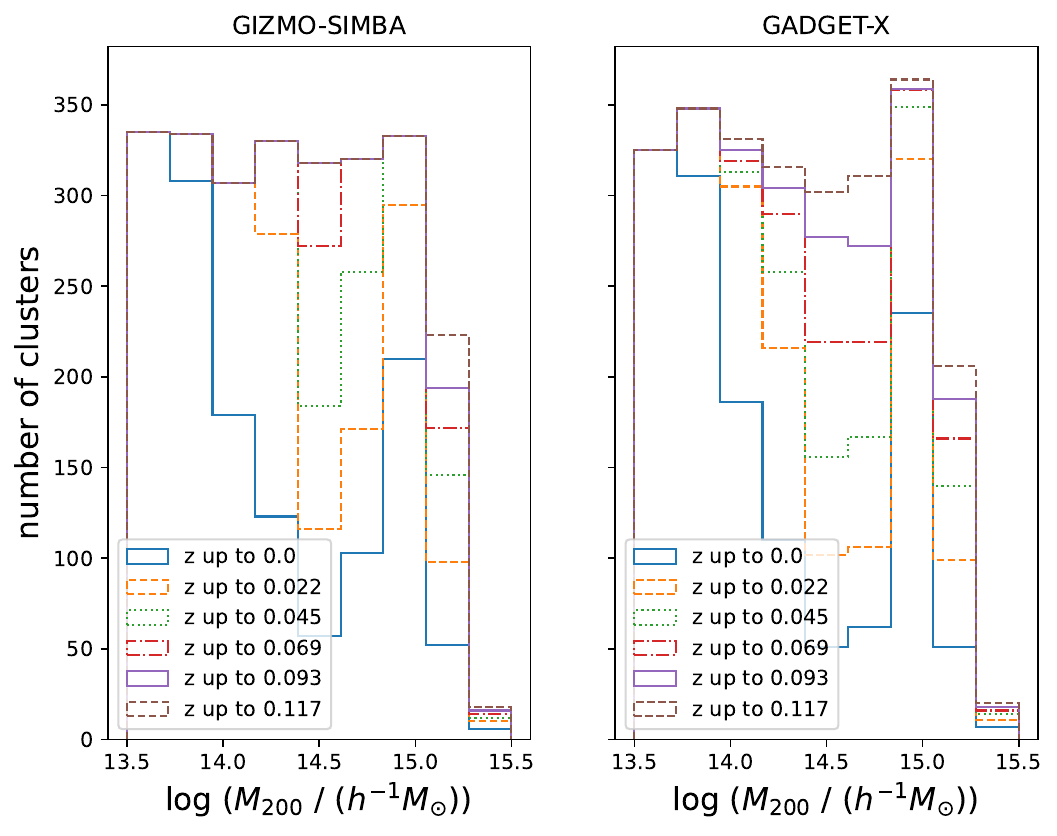}
\caption{Number of selected clusters as a function of mass for our two simulations \simba{} and \gadgetx{}. The sample is increased when not only snapshots at z=0.0 are considered and objects up to z=0.117 are selected.}
\label{fig:Mdistsimulations}
\end{figure}

\begin{figure}
\includegraphics[width=\columnwidth]{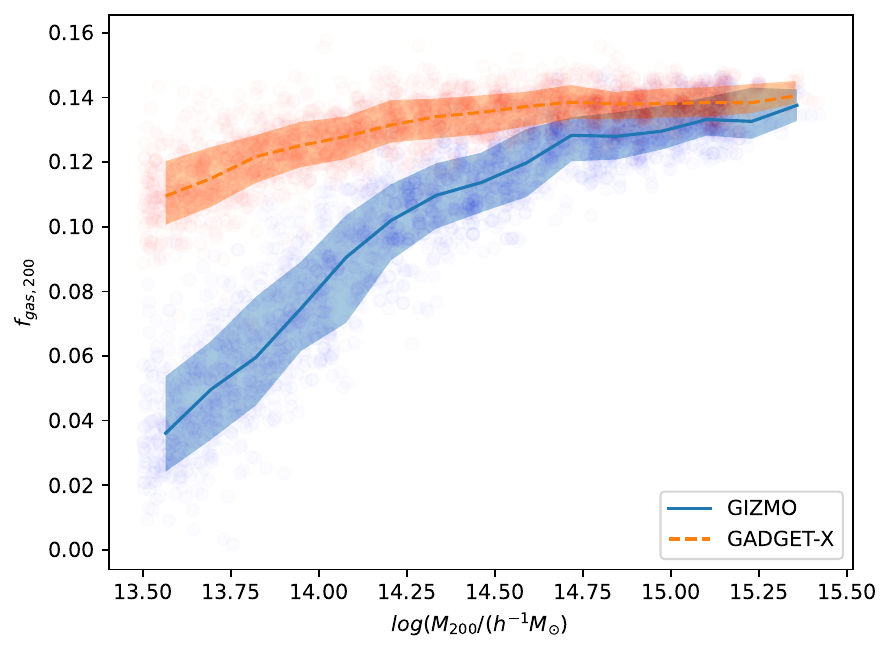}
\caption{Median gas fraction as a function of the mass for the \simba{} and GADGET-X simulations. The shaded regions correspond to the $16^{\text{th}}$ and $84^{\text{th}}$ percentiles.}
\label{fig:fgas}
\end{figure}
\begin{figure}
\includegraphics[width=\columnwidth]{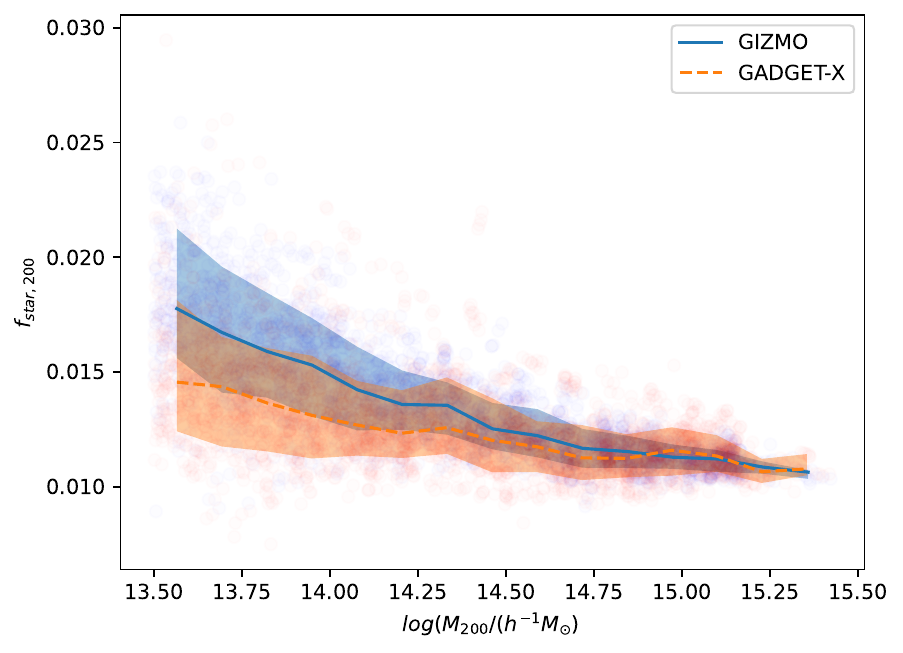}
\caption{Median star fraction as a function of the mass for the \simba{} and GADGET-X simulations. The shaded regions correspond to the $16^{\text{th}}$ and $84^{\text{th}}$ percentiles.}
\label{fig:fstar}
\end{figure}
\begin{figure*}
\includegraphics[width=\textwidth]{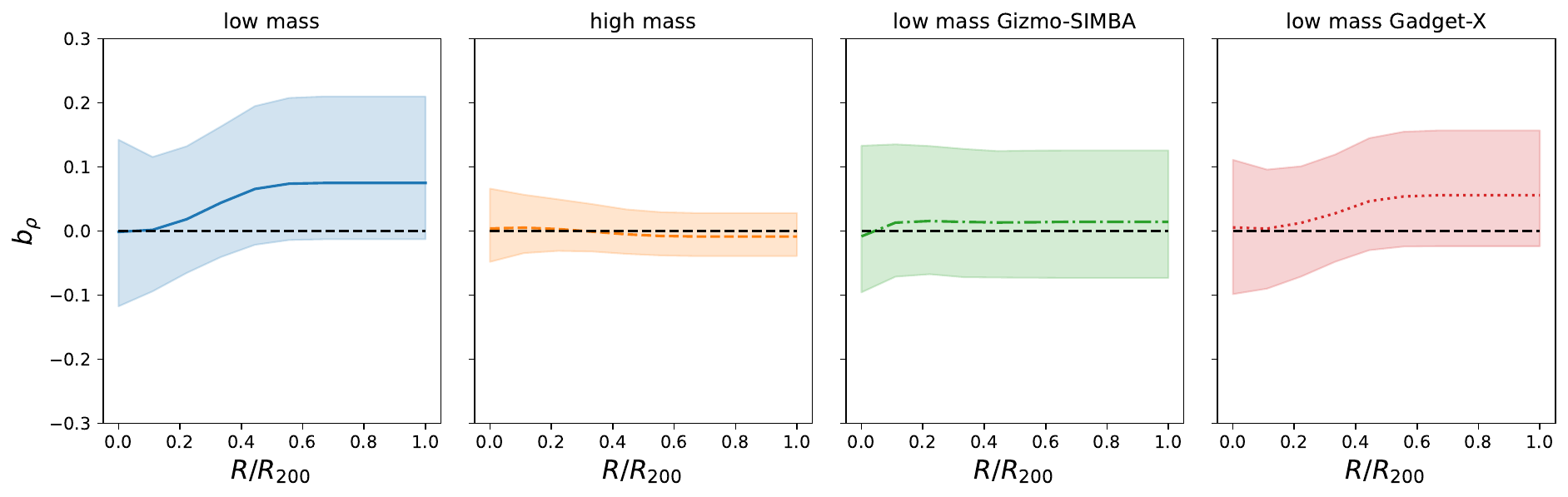}
\caption{Mass bias see Eq.(\ref{eq:bias}) for the mass profiles as a function of the radius $R/R_{200}$ for the experiments previously mentioned in this appendix. From left to right, we show the bias corresponding to the different experiments: Train with low-mass objects with data from both simulations, train only with high-mass objects with data from both simulations, train low-mass objects with data only from \simba{} and train with low-mass objects with data only from \gadgetx{}. The lines represent the medium values per bin and the shaded regions cover the $16^{\text{th}}$ and $84^{\text{th}}$ percentiles.}\label{fig:bias_appendx}. 
\end{figure*}

Throughout this work, we have shown that the difference between the ground-truth maps from the predicted maps is dependent on the cluster mass $M_{200}$ (see \autoref{fig:mass_profiles} and \autoref{fig:MMDs}). We have claimed that this trend is primarily due to the fact that gravity is more relevant for massive galaxy clusters and astrophysical effects implemented in {\theth{}} simulations, which are different codes for star formation, supernovae and black hole feedbacks, become more relevant in small groups and thus, the scatter of the predicted mass density maps of these small objects increases.

Furthermore, as mentioned in \autoref{sec-2}, we have trained with data from \gadgetx{} and \simba{} simulations and the CNN model has managed to learn from both simulations simultaneously. The particular distribution of selected objects is shown in \autoref{fig:Mdistsimulations}. However, both codes generate very different types of galaxy clusters as far as the ICM and star properties are concerned. In \autoref{fig:fgas}, the gas fraction of our data sample is shown for \simba{} (blue) and \gadgetx{} (dashed orange). The figure implies that the gas fraction for \gadgetx{} is more constant, while for \simba{} decays faster in small groups. Nevertheless, the star fractions follow the opposite trend, as shown in \autoref{fig:fstar}. In summary, these two figures show that different astrophysical implementations converge in the massive end and therefore, one might expect that the ICM and star distributions follow gravity, and are more related to the total mass density of the galaxy clusters for massive galaxy clusters.

To show that the scatter does not depend on the machine learning model, but rather on the diversity of the data, we have run different experiments:
\begin{enumerate}
    \item \textbf{Low mass:} We have retrained the multi-3 model with galaxy clusters belonging only to the first mass interval, i.e., $13.5\leq\log M_{200}/\hMsun<14.02$.
    \item \textbf{High mass:} We have also retrained the model with data belonging only to the most massive interval ($14.63\leq\log M_{200}/\hMsun\leq15.38$). 
    \item \textbf{Low mass \simba{}:} We have retrained the model multi-3 with data belonging only to \simba{} only in the first mass interval.
    \item \textbf{Low mass Gadget-X:} We have also retrained the multi-3 model with data belonging to \gadgetx{} only in the first mass interval.
\end{enumerate}
The results of the experiments are shown in the \autoref{fig:bias_appendx}. From the first and second experiments, we conclude that the scatter does not depend on the ML model and training procedure. We acknowledge that although these could bias the prediction to be more accurate at higher masses, that is not the case as the experiments suggest. From the third and fourth experiments, the results on \simba{} and GADGET-X simulations are similar and therefore, we conclude that the performance should not vary significantly for different hydrodynamic simulations.


\bsp	
\label{lastpage}
\end{document}